\documentclass[10 pt, thmsa,letterpaper]{article}
\usepackage{sw20lart}

\input tcilatex
\QQQ{Language}{
American English
}

\begin{document}

\author{NEACSU MARIA CRISTINA \\
%EndAName
Department of Physics\\
Technical University ''Gh. Asachi'' \\
Str. Copou Nr.22, Iasi, 6600\\
ROMANIA\\
e-mail:mneacsu@mt.tuiasi.ro}
\title{DILATION DARK MATTER IN\ VAIDYA-DE-SITTER SPACETIME }
\maketitle

\begin{abstract}
The exterior of a relativistic star can be modelated with the Vaidya
radiating metric. It is started from the generalized Vaidya metric that
allows a type II fluid and studied the conditions of generating new
analytical solutions of the Einstein's field equations. It is shown that the
mass parameter solution gives the classical de Sitter universe in the static
case and the extended de Sitter metric coupled with a dilation scalar field
in the time-dependent case. It is concluded that in the time-dependent case
the atmosphere of a relativistic star consists on an anisotropic string
fluid coupled with a dark matter null fluid and interpreted the scalar field
as the particle that produces the dark matter.
\end{abstract}

\section{Introduction}

The exterior of relativistic stars allows atmosphere to be added to
classical vacuums \cite{1} and many classical vacuum models may have
atmosphere composed of a fluid or field of strings \cite{2}. Strings played
a role in the seeding of density inhomogeneities in the early universe \cite
{3}, \cite{4}.

Recently was pointed out that allowing the Schwarzschild solution mass
parameter to be a function of radial position, creates an atmosphere with a
string fluid stress-energy around a static, spherically symmetric object. If
the mass is also a function of retardate time, we find the Vaidya metric of
a short-wavelength photons radiation atmosphere in addition to the string
fluid \cite{5}.

We will start from the generalized Vaidya solution, which includes most of
the known solutions of the Einstein field equations, such as: the monopole,
the de Sitter, the charged Vaidya and the Husian solution The generalization
is possible because the energy-momentum tensor is linear in terms of the
mass function. Since the metric has one arbitrary function: $m(u,r)$, a
given mass distribution determines the stress-energy and the linear
superposition of particular solutions is also a solution of the Einstein
field equation \cite{6}.

In this work it is studied the extension of Vaidya's metric with arbitrary
mass function, which allows an atmosphere of a string fluid and an
associated fluid. It is obtained analytic mass solutions that leads to the
de Sitter metric for the static case and for the extended de Sitter metric
coupled with a dilation scalar field in the time-dependent case. It is shown
that the second case allows the atmosphere of the relativistic star to have
a string fluid and a black matter fluid. The metric signature is $-2$, latin
indices range will go over the coordinates ($u,r,\theta ,\varphi $ )
overdots abbreviate $\partial /\partial _t$ and primes $\partial /\partial _r
$; overhead denotes unit vectors and the constants are: $G=c=1$.

\section{The generalized Vaidya solution}

The spacetime covering the exterior of a spherical star is given by: 
\begin{equation}
ds^2=a\,du^2-2\,\varepsilon \,du\,dr-r^2\left( d\theta ^2+\sin ^2\theta
\,d\varphi ^2\right) \text{,}  \tag{1}
\end{equation}
where $a=1-2m(u,r)/r$ and $m(u,r)$ is the mass function. When $\varepsilon
=+1$ the null coordinate $u$ represents the Eddington advanced time, in
which $r$ is decreasing towards the future along the ray $u=const$
(ingoing), and $\varepsilon =-1$ represents the Eddington retardate time, in
which $r$ is increasing (outgoing). We will consider $\varepsilon =-1$ that
gives the metric in retardate time.

The Einstein tensor is computed using the GRTensor package, within the
Newman-Penrose null tetrad formalism and nonholonomic bases. The metric (1)
has the principal null geodesic vectors $l_a,n_a$ of the form: 
\begin{equation}
l_a=\delta _a^u,\qquad n_a=a/2\delta _a^u+\delta _a^r,  \tag{2}
\end{equation}
where: $l_al^a=n_an^a=0,$ $l_an^a=-1$ and admits an ortonormal basis defined
by the following four unit vectors: 
\begin{equation}
\hat{u}_a=a^{1/2}\delta _a^t+a^{-1/2}\delta _a^r,\qquad \hat{r}%
_a=a^{-1/2}\delta _a^r,\qquad \hat{\theta}_a=r\delta _a^\theta ,\qquad \hat{%
\varphi}_a=r\sin \theta \delta _a^\varphi ,  \tag{4}
\end{equation}
\begin{equation}
\tag{5}
\end{equation}
The Einstein tensor reads: 
\begin{equation}
G_{ab}=\frac{2\dot{m}}{r^2}l_al_b-\frac{2m^{^{\prime }}}{r^2}\left( \hat{u}_a%
\hat{u}_b-\hat{r}_a\hat{r}_b\right) +\frac{m^{^{\prime \prime }}}r\left( 
\hat{\theta}_a\hat{\theta}_b+\hat{\varphi}_a\hat{\varphi}_b\right) .  \tag{6}
\end{equation}
Now, using the Einstein field equations: $G_{ab}=-8\pi T_{ab\text{ }}$, it
is seen that the energy-momentum tensor can be written as two-fluid system: $%
T_{ab}=$ $T_{ab}^{(n)}+$ $T_{ab}^{(m)}$ , where: 
\begin{equation}
T_{ab}^{(n)}=\psi l_al_b  \tag{7}
\end{equation}
it is the null fluid tensor corresponding to the component of the matter
field that moves along the null hypersurfaces $u=const.$, and: 
\begin{equation}
T_{ab}^{(m)}=\left( \rho +p\right) \left( l_an_b+n_al_b\right) +pg_{ab} 
\tag{8}
\end{equation}
is the matter fluid. From the identification on obtain the following
expression for the matter and radiation quantities: 
\begin{equation}
\psi =\frac{\dot{m}(u,r)}{4\pi r^2},\qquad \rho =\frac{m^{\prime }(u,r)}{%
4\pi r^2},\qquad \mu =\frac{m^{^{\prime \prime }}(u,r)}{8\pi r}.  \tag{9}
\end{equation}
This identification leads to an equation of state: 
\begin{equation}
\rho +p=0.  \tag{10}
\end{equation}

The restriction (10) is respected for an energy-momentum tensor that
describes the string fluid situation \cite{2}. The string bivector is
defined by \cite{3}: 
\begin{equation}
\sigma ^{ab}=\eta ^{\alpha \beta }\frac{\partial x^a}{\partial x^\alpha }%
\frac{\partial x^b}{\partial x^\beta }.  \tag{11}
\end{equation}
Spherical symmetry demands that the averaged string bivector will have a
word-sheet in the $(u,r)$ plane. In terms of unit vectors $\sigma ^{ab}$
will be: 
\begin{equation}
\sigma ^{ab}=\hat{r}^a\hat{u}^b-\hat{u}^a\hat{r}^b  \tag{12}
\end{equation}
and the condition $\gamma =1/2\sigma ^{ab}\sigma _{ab}$ implies that only
the $\sigma ^{ur}$ component is non-zero, therefore on obtain: 
\begin{equation}
\sigma ^{ac}\sigma _c^b=\hat{u}^a\hat{u}^b-\hat{r}^a\hat{r}^b.  \tag{13}
\end{equation}

The string energy-momentum tensor, by analogy with the one for the perfect
fluid, is written as \cite{4}: 
\begin{equation}
T_{ab}^{\left( s\right) }=\rho \sigma _a^c\sigma _{cb}-p_{\bot }h_{ab}, 
\tag{14}
\end{equation}
where $\hat{h}_a^b=\delta _a^b-\sigma _{ac}\sigma ^{cb}$ and $\hat{h}%
_c^a\sigma ^{cb}=0$. The restriction (10) holds because the energy-momentum
tensor (14) can be written in the form: $T_{ab}^{}=\rho u_au_b-px_ax_b$ and
the energy condition $g\left( x_\alpha ,x^\alpha \right) =1$, where $u_a=%
\hat{u}_a$ and $x^\alpha =\left( \hat{r},\hat{\theta},\hat{\varphi}\right) $%
, is respected. We thus verified that the  energy-momentum tensor do not
describe the tachionic matter, but the string fluid matter.

In terms of the unit vectors, the string fluid becomes: 
\begin{equation}
T_{ab}^{(s)}=\rho \hat{u}_a\hat{u}_b+p_r\hat{r}_a\hat{r}_b+p_{\bot }\left( 
\hat{\theta}_a\hat{\theta}_b-\hat{\varphi}_a\hat{\varphi}_b\right)   \tag{15}
\end{equation}
The expression (15) can be identified, through Einstein field equations,
with eq. (6) for the Einstein tensor and for the physical quantities on
obtain: 
\begin{equation}
\psi =-\frac{\dot{m}}{r^2},\qquad \rho =-p_r=\frac{m^{^{\prime }}}{4\pi r^2}%
,\qquad p_{\bot }=-\frac{m^{^{\prime \prime }}}{8\pi r}.  \tag{16}
\end{equation}
Those quantities describe a type II fluid, composed with a string fluid with
equation of state $\rho +p_r=0$ and radial pressure equal with transverse
stress $p_r$ $=p_{\bot }$ ; coupled with radiation.

\section{Analytic mass solutions}

Without loss of generality, $m(u,r)$ can be expanded in powers of $r$ ,
named: 
\begin{equation}
m(u,r)=\stackrel{+\infty }{\stackunder{n=-\infty }{\sum }}a_n(u)r^n, 
\tag{17}
\end{equation}
where $a_n(u)$ are arbitrary functions of $u$ only. The substitution of (17)
into (16) yields: 
\begin{equation}
\begin{array}{c}
\psi =\frac 1{4\pi }\stackrel{+\infty }{\stackunder{n=-\infty }{\sum }}\dot{a%
}_n(u)r^{n-2},\quad \rho =\frac 1{4\pi }\stackrel{+\infty }{\stackunder{%
n=-\infty }{\sum }}na_n(u)r^{n-3}, \\ 
p_{\bot }=-\frac 1{8\pi }\stackrel{+\infty }{\stackunder{n=-\infty }{\sum }}%
n(n-1)a_n(u)r^{n-3},
\end{array}
\tag{18}
\end{equation}
The above solutions include a large class of the Einstein field equations
with spherical symmetry.

\subsection{The static case}

It is considered first the static model, when the mass function is
time-independent and $a_n(u)$ is an arbitrary constant. The density is
time-independent too and on find the situation of a static, isotropic string
fluid cloud coupled with a null radiative fluid. The mass is determined from
(16) by the constraint $p_r=p_{\bot }$, which yields: 
\begin{equation}
m(r)=a_1+a_3r^3.  \tag{19}
\end{equation}
The $-\rho =p_r=p_{\bot }$ restriction imposes in (18) the maximum $n=3$ for
the sum index. After the substitution of (19) into (18) and the expansion of
(18) for $n=3$, it is found: 
\begin{equation}
\psi =0,\quad \rho =\frac 1{4\pi }\left( \frac{a_1}{r^2}+3a_3\right) ,\quad
p_{\bot }=-\frac{3a_3}{4\pi }.  \tag{20}
\end{equation}
Now, the same restriction is imposed, which implies that the constant $a_1=0$
and: 
\begin{equation}
a_3=\frac{4\pi \rho }3\Longrightarrow m(r)=\frac{4\pi r^3}3\rho   \tag{21}
\end{equation}
Can be made the identification $a_3=\Lambda /6$. In this way, it is found
the classical vacuum of the de Sitter spacetime described by the
energy-momentum tensor $T_{ab}=-\Lambda g_{ab}$ with vanishing null
radiation fluid and cosmological constant $\Lambda =8\pi \rho .$

\subsection{The time-dependent case}

In this case the radiation null fluid is added to the string fluid and
yields a two-fluid atmosphere of an anisotropic string fluid cloud in
interaction with the null radiative fluid. The solution (19) in the more
general time-dependent case is: 
\begin{equation}
m(u,r)=a_1(u)+a_3(u)r^3.  \tag{22}
\end{equation}
Then (18) becomes: 
\begin{equation}
\psi =-\frac 1{4\pi }\frac{\dot{a}_3(u)}r,\quad \rho =\frac 3{4\pi }a_3(u). 
\tag{23}
\end{equation}
When on integrate, on obtain the solution: 
\begin{equation}
a_3=e^{-\frac 3r\int \frac \psi \rho du}  \tag{24}
\end{equation}
From the eq. (16), the integrability condition for $m$ is written as: 
\begin{equation}
\dot{\rho}+\psi ^{^{\prime }}+\psi /r=0  \tag{25}
\end{equation}
This is a general form and any particular form for $\psi $ and $\rho $ will
give a solution and those functionsadmit the expansion in powers of $r$,
without loss of generality. It is considered the case when the string
density is $\rho =\rho _s+kt$, where $\rho _s$ is time-independent and $%
k=const$. Then eq. (25) becomes: 
\begin{equation}
\psi ^{^{\prime }}+\psi /r=k  \tag{26}
\end{equation}

In the measurements of the dark matter localized on the equator of the
galaxy \cite{7}, it was used an energy-density profile for dark matter of
the form: 
\begin{equation}
\psi (r)=\frac{\rho _0r_c^2}{r^2+r_c^2}  \tag{27}
\end{equation}
It is considered the constant $r_c\rightarrow 0$ and the retardate time
coordinate $u\rightarrow t$. Then (27) becomes $\psi (r)=\rho _0/r^2$ that
can be substituted in (25) and on obtain a solution for the string fluid
density: $\rho =\rho _s+\rho _0t/r^3$. Any expansion of anisotropic $\rho _s$
in terms of $r$ can be a new solution. It can be taken $\rho _s=0$ , then
the eq. (24) is written in this form: 
\begin{equation}
a_3=te^{-3}.  \tag{28}
\end{equation}

\section{Discussions}

The equation (24) is a very interesting result, because considering the case
of the de Sitter solution again, the exponential term appears as an
extension of the cosmological constant: 
\begin{equation}
\Lambda \rightarrow e^{-\frac 3r\int \frac \psi \rho du}\Lambda .  \tag{29}
\end{equation}
This is similar with that one obtained from the low energy limit of
Superstring theory. The general Lagrangian obtained from high-dimensional
theories, after dimensional reduction and conformal transformations, has the
form: 
\begin{equation}
L_4=-R_4+2(\nabla \Phi )^2+e^{-2\alpha \Phi }\Lambda ,  \tag{30}
\end{equation}
where $\Phi $ is the scalar field coupled with matter and $\alpha $ is the
coupling constant.

Let take the Kaluza-Klein theories coupling constant that is $\alpha ^2=3$
and the radiation null fluid being the dark matter fluid. Then on compare
(28) with the third term of (30) and on obtain for the scalar field the
expression: 
\begin{equation}
\Phi =\frac{\sqrt{3}}2\ln t.  \tag{31}
\end{equation}
The scalar field is time dependent and determines the time expansion of the
universe, i.e., has the role of a dilation field. The general form of
eq.(31), when the string density is a function $\rho =\rho (t)/r^3$, yields: 
\begin{equation}
\Phi =\frac 3{2\alpha }\rho _0\int \frac{dt}{\rho (t)}.  \tag{32}
\end{equation}

In this case, the scalar field keeps the information about the dark matter
and it can be interpreted as a field that introduces a time energy of
radiative type, i.e., the dark matter is produced by a $\Phi $ particle.
This is in accord with the most recent discoveries that supports the
existence of a anti-gravity force that counteract the gravitational forces
and is causing the expansion of the universe to accelerate. Scientists
referred this force as cosmic dark energy, revived in the cosmological
constant and describe this cosmic dark energy as ''a vacuum energy assigned
to empty space itself, a form of energy with negative pressure.''
Understanding the source and nature of this force gives scientists a
radically new picture of the future of the universe. It appears that the
dark energy could eventually overwhelm the gravitational forces of matter.
The density of matter in the universe would then become insignificant, so
that the universe would approach an essentially uniform force field of dark
energy.

\end{document}